**[Article Full Title]**

# Generalizable dose denoising using diffusion transformer framework for pencil beam scanning proton therapy

**[Short Running Title]**

Generalizable dose denoising in PBSPT

**[Author Names]**

Yuzhen Ding[1], Hongying Feng[1], Jianfeng Zhao[1], Jason Holmes[1], Martin Bues[1], Lisa A. McGee[1], Jean-Claude M. Rwigema[1], Nathan Y. Yu[1], Terence T. Sio[1], Sameer R. Keole[1], William W. Wong[1], Steven E. Schild[1], Jonathan B. Ashman[1], Sujay A. Vora[1], Daniel J. Ma[2], Samir H. Patel[1], and Wei Liu[1]

**[Author Institutions]**

[1]Department of Radiation Oncology, Mayo Clinic, Phoenix, AZ 85054, USA

[2]Department of Radiation Oncology, Mayo Clinic, Rochester, MN 55905, USA

**[Corresponding Author Name & Email Address]**

Wei Liu, PhD. Department of Radiation Oncology, Mayo Clinic Arizona, 5777 E. Mayo Blvd. Phoenix, AZ 85054. Telephone: 480-342-4215. Fax: 480-342-3972. E-mail: Liu.Wei@mayo.edu.

**[Data Availability Statement for this Work]**

Research data are stored in an institutional repository and will be shared upon request to the corresponding author. Owing to institutional regulations and patient privacy considerations, the complete training dataset cannot be made publicly available. The implementation code released in GitHub https://github.com/Mayo-Clinic-RadOnc-Foundation-Models/DiT-dose-denoising-in-PBSPT with DOI: 10.5281/zenodo.23027840.

**[Acknowledgements]**

None

**[Key Words]**

Dose denoising, diffusion transformer, Monte Carlo, intensity-modulated proton therapy

**[Word Count]**

Abstract: 247; manuscript: 2997

**Abstract**

**Background and purpose:** Intensity-modulated proton therapy (IMPT) enables conformal dose delivery for head and neck (H&N) cancers. However, its sensitivity to inter-fractional anatomical changes motivates online adaptations, where rapid and accurate Monte Carlo (MC) dose calculation remains a major bottleneck. Low-statistics MC simulations reduce computation time but introduce noise. This study aimed to develop a diffusion-transformer framework to recover high-statistics dose distributions from noisy MC simulations.

**Materials and methods**: A diffusion-transformer-based dose denoising framework was developed. Eighty H&N patients were used for training and additional 10 H&N patients and 30 patients with lung, breast, and prostate cancers (10 per site) were used for evaluation. Noisy and high-statistics dose maps were generated using MCsquare. The model was trained using noisy dose maps and CT images as inputs and high-statistics dose maps as ground truth, with mean squared error, residual, regional mean absolute error (MAE), and dose-volume histogram (DVH) losses. Performance was evaluated using MAE, 3D gamma passing rates, and DVH indices.

**Results:** Mean whole-body MAE ranged from 0.08 ± 0.06 to 0.20 ± 0.11 Gy (Relative Biological Effectiveness, RBE) (RBE = 1.1) across test cohorts. Mean whole-body 3 %/2 mm gamma passing rates ranged from 95.8 ± 0.8 % to 97.6 ± 0.9 %. DVH indices from high-statistics and predicted dose distributions showed close agreement for clinical target volumes and organs of interest.

**Conclusions**: The diffusion-transformer-based framework denoised low-statistics MC dose distributions. Despite being trained on H&N data, the model generalized to lung, breast, and prostate cohorts without retraining.

## 1. Introduction

Intensity-modulated proton therapy (IMPT) leverages the Bragg peak to deliver highly conformal dose distributions with minimal exit dose, enabling superior sparing of normal tissues compared with photon therapy [1-3]. This advantage is particularly important for head and neck (H&N) cancers, where targets are adjacent to numerous critical organs of interest (OoIs), contributing to improved outcomes regarding adverse events, quality of life and survival [4-6].

Nevertheless, IMPT is highly sensitive to range and setup uncertainties [7-18] and to intra-fractional anatomical motion, which can substantially degrade plan quality [19-26]. Online adaptive radiation therapy (oART) has emerged to enable plan adaptation to daily anatomic variations. A major barrier to clinical implementation of proton oART is fast and accurate dose calculation. Monte Carlo (MC) [27-31] simulations, while considered the gold standard, still impose substantial computational burdens in time-sensitive workflows, particularly when repeated dose recalculations and robustness evaluations are required. In contrast, pencil beam algorithms (PBA) [32-35] are computationally efficient but less accurate in heterogeneous tissues. Thus, neither approach is ideal for time-sensitive proton oART workflows.

Recent advances in artificial intelligence and deep learning in radiotherapy [36, 37] have enabled data-driven dose prediction and denoising models that approximate MC accuracy with substantially reduced computation time. Several studies have demonstrated deep learning–based dose prediction in photon and proton therapy using U-Net-based patient-specific models, knowledge-informed neural networks, hybrid convolutional and transformer architectures, anatomy- and feature-informed neural networks, and beam- and plan-conditioned dose-prediction models [38-44]. An alternative strategy is deep learning–based MC dose denoising, which reconstructs high-statistics dose distributions from faster low-statistics MC simulations while retaining patient- and plan-specific physical dose information [45,46]. Many direct dose-prediction approaches are disease-site specific and rely on extensive anatomy-, and plan-specific inputs, including clinical target volume (CTV) and OoI contours, computed tomography (CT)-derived features, beam-path information, spot energies and weights distributions, and other customized planning information [38–44]. Proton MC denoising reduces this input burden by starting from a physics-based dose distribution, but prior studies have not systematically evaluated transfer from a single heterogeneous disease site to previously unseen sites [45,46]. Robustness and generalizability across heterogeneous clinical conditions therefore remain incompletely characterized. These limitations may hinder clinical translation, particularly for highly heterogeneous sites such as H&N and for adaptive workflows involving domain shifts across disease sites, prescription doses, beam configurations, and imaging conditions.

In this study, we developed a diffusion-transformer (DiT)-based dose denoising framework using limited and loosely constrained input data. No restrictions were imposed on sub-disease categories, treatment plan configurations, CT image resolutions, or other patient-specific variables to better reflect clinical practice. We investigated whether a diffusion-transformer framework trained on heterogeneous H&N data could generalize to additional disease sites under clinically realistic conditions. The framework was trained and evaluated using data from 90 H&N patients and further tested on 30 patients with lung, breast, and prostate cancers to provide an initial assessment of cross-site generalizability.

# 2. Materials and methods

## 2.1 Data collection and processing

Following institutional review board approval, we collected treatment plans and CT images from 90 patients with H&N cancer who received IMPT at our clinic. Using the open-source MC software, MCsquare [31], two dose distributions were generated for each patient: (1) a high-statistics reference distribution and (2) a low-statistics dose distribution generated using one-hundredth of the particles employed for the high-statistics simulation. All dose distributions were calculated using a dose-grid resolution of $2.5 \times 2.5 \times 2.5$ mm$^3$.

No additional selection criteria, including sub-disease site, age, sex, or treatment history, were imposed to preserve dataset heterogeneity. Consequently, substantial inter-patient variations in CT images and dose distributions were present.

To standardize the input data while preserving patient-specific variation, the following preprocessing pipeline was developed (Figure S1 in Supplementary Material A): 1) The low-statistics and corresponding high-statistics dose distributions were vectorized into one-dimensional arrays. 2) The dose vectors were divided into non-overlapping chunks of size $1\times4096$ with zero-padding and reshaped to $4 \times 32 \times 32$. Chunks containing only zero-dose voxels were discarded. 3) A Gaussian filter with $\sigma = 2.5$ mm was applied to the low-statistics dose distributions as a fixed preprocessing step. 4) All dose chunks were normalized to the range [0,1]. During training, the paired low- and high-statistics dose distributions for each patient were normalized by the same scalar, defined as the corresponding maximum high-statistics dose (Gy (Relative Biological Effectiveness, RBE)). A constant proton RBE of 1.1 was used. This scalar was used only for numerical scaling and was not provided to the network as an additional predictive input. During inference, because the high-statistics reference dose was unavailable, the clinically available prescription dose (Gy (RBE)) was used for normalization and subsequent rescaling to the physical dose domain. Subsequently, 1 was added to each normalized voxel value and a logarithmic transformation was applied, thereby compressing the dynamic range of the dose values. 5) CT images were vectorized, divided into chunks of size 1×4096, and reshaped to 4×32 ×32, with those corresponding to the removed dose chunks also excluded. 6) CT Hounsfield unit (HU) values were nonlinearly mapped to the range [0,1], with emphasis on the interval from −200 to 300 HU, where most soft tissues and tumors are located. Voxels with HU values below −1000 and above 29,000 were clipped. The mapping function is summarized in Table S1 in Supplementary Material A. 7) To improve superior–inferior continuity and increase sample diversity, both dose distributions and CT images were transposed and steps 1–6 were repeated. During inference, denoised chunks were transformed back to the original scale and spatially concatenated to reconstruct the final dose distribution.

Of the 90 patients, 80 were used for training and 10 were reserved for internal testing. Structure contours were not provided to the network as predictive or conditioning inputs. Following preprocessing, the training set comprised 124,493 triplets of low-statistics dose, high-statistics dose, and corresponding CT images, where the low-statistics dose and CT images served as network inputs and the high-statistics dose served as the ground truth.

## 2.2 Diffusion transformer-based dose denoising framework

We developed a dose denoising framework based on the DiT [47] architecture, originally designed for high-fidelity image generation. The model was trained to recover high-statistics MC dose distributions from low-statistics noisy dose maps using a conditional diffusion formulation (Figure 1).

Unlike standard DiT implementations, the pretrained variational autoencoder was removed due to its incompatibility with ultra-sparse dose distributions. Instead, the diffusion model was directly conditioned on the noisy dose map and the corresponding CT image to incorporate anatomical and dose distribution information. To focus on learning clinically relevant regions, CT images were masked to the non-zero dose region and used as conditioning inputs.

The diffusion model was trained using a noise-prediction objective combined with additional clinically motivated losses, including a regional mean absolute error (MAE) emphasizing high-dose voxels, a residual loss targeting low- and high-dose extremes to preserve Bragg peak characteristics, and a dose-volume histogram (DVH)-based loss [48] to capture clinically relevant dose metrics. The total loss was defined as the weighted sum of these components. Detailed formulations of the diffusion process, conditioning strategy, loss functions, and training procedure are provided in the Supplementary Materials.

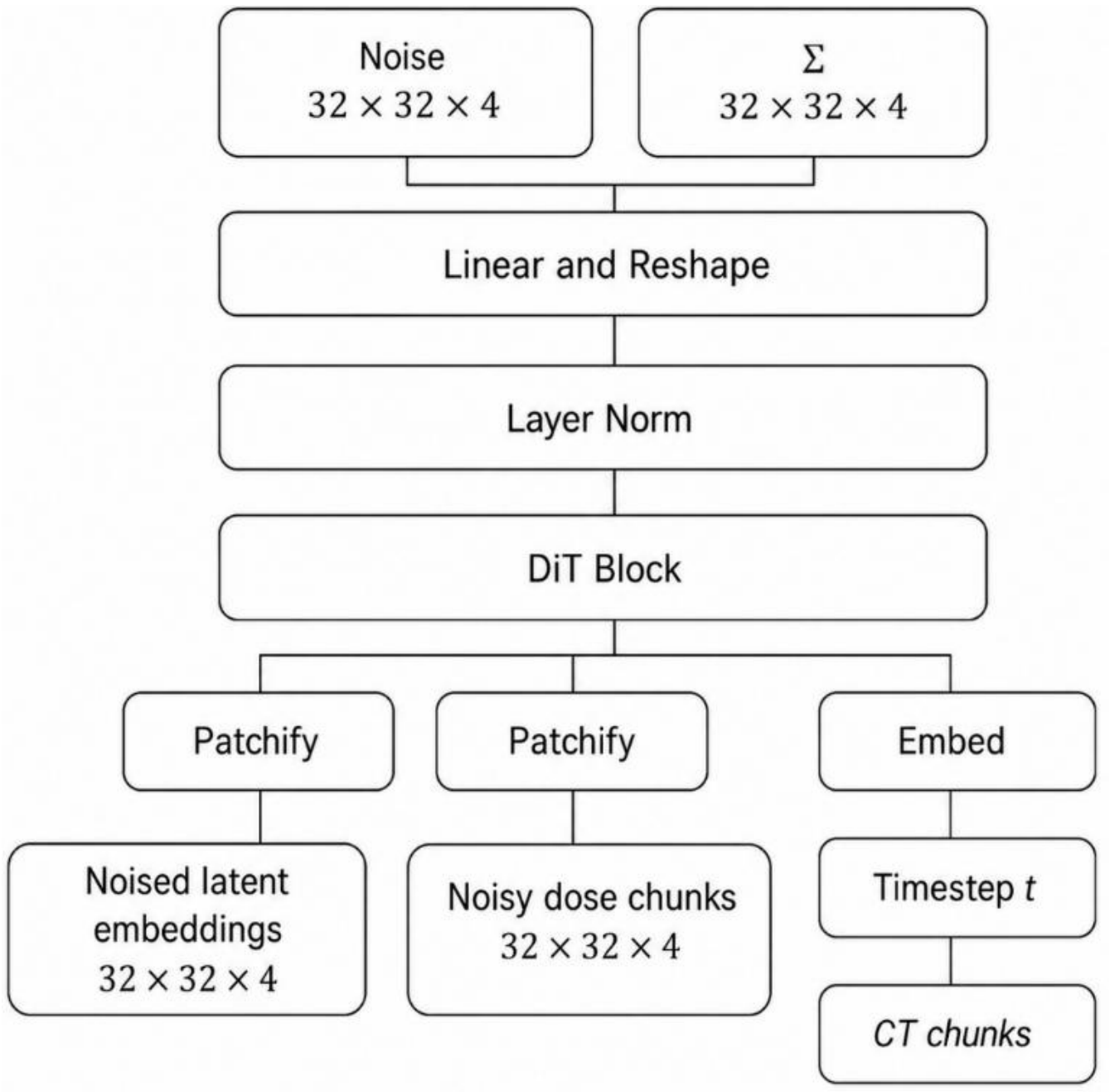

**Figure 1.** Model architecture. The low-statistics dose map and diffusion-noised latent embeddings derived from the high-statistics dose map were used as model inputs, while the corresponding CT images served as conditioning inputs. The proposed framework consisted of eight DiT blocks employing standard transformer blocks with adaptive layer normalization.

### 2.3 Model evaluation and statistical analysis

To quantitatively assess the discrepancies between the predicted dose map $\hat{x}_0$ and ground truth dose map $x_0$, MAE was calculated for the 10 H&N patients reserved for testing. DVH indices were calculated to visualize differences between the predicted and ground-truth dose maps. In addition, with special emphasis on treatment plan quality, we also calculated CTV $D_{98\,\%}$ (%) and CTV $D_{2\,\%}$ (%) to evaluate target dose conformality and homogeneity, respectively. Here $D_{x\,\%}$ (%) is defined as the minimum dose (normalized by the prescription dose) that covers at least $x$ % CTV volume. Clinically relevant DVH indices for OoIs were also reported across all four disease sites. Global 3D gamma analysis was performed between the denoised and high-statistics reference dose distributions using a 3 %/2 mm criterion, with a 10 % low-dose threshold and a global normalization to the maximum dose of the reference distribution. The reference dose distribution was sampled at 0.5 mm spatial increments using trilinear interpolation. Because the native MC dose grid was $2.5 \times 2.5 \times 2.5$ mm$^3$, the 3 %/2 mm criterion was used as a clinically familiar reference benchmark rather than as a universal standard for evaluating dose-denoising models. A global 2 %/2 mm analysis was additionally performed as a sensitivity analysis and is reported in Supplementary Materials. Furthermore, to quantitatively assess generalizability, evaluations were repeated on an independent cohort of 30 patients across prostate, breast, and lung cancers (10 per site).

The Gaussian filter applied to the low-statistics dose ($\sigma = 2.5$ mm) was optimized and evaluated as a standalone denoising baseline. Fifteen head-and-neck patients were randomly drawn from the training cohort and used exclusively to select σ, with whole-body MAE in Gy (RBE) as the single prespecified primary metric; no test-cohort patient contributed to the selection. Six filter widths ($\sigma = 1.25$–$7.50$ mm) were evaluated, and $\sigma = 1.88$ mm yielded the lowest validation MAE and was therefore applied to all test cohorts without further tuning. All baseline and proposed-method evaluations used identical test cases, structure masks, normalization, and 3D gamma-analysis.

Model training and inference were performed using Python (version 3.9.7) and PyTorch (version 2.7.1) with Compute Unified Device Architecture (CUDA; version 12.6) acceleration on an NVIDIA A100-SXM4-80GB graphics processing unit (GPU; NVIDIA, Santa Clara, CA, USA). Statistical analyses and data visualization were performed using the NumPy (version 1.26.4), SciPy (version 1.10.1), and Matplotlib (version 3.5.3) Python libraries.

## 3. Results

Across the four testing cohorts, MAE values ranged from 0.08 ± 0.06 to 0.20 ± 0.11 Gy (RBE) for the whole body, from 0.14 ± 0.36 to 1.7 ± 1.6 Gy (RBE) for the CTV, and from 0.11 ± 0.07 to 0.33 ± 0.31 Gy (RBE) for OoIs (Table 1). Representative dose-difference maps showed spatially localized differences between the denoised and high-statistics reference dose distributions in H&N, lung, breast, and prostate cases (Figure 2).

Differences in CTV DVH indices between the denoised and high-statistics reference dose distributions were 1.92 ± 0.03 % and 1.8 ± 1.1 % for $D_{98\,\%}$ and $D_{2\,\%}$, respectively. OoI DVH indices also showed limited numerical differences between the denoised and reference dose distributions across the evaluated disease sites. Representative DVHs showed close overlap between the denoised and high-statistics reference dose distributions across H&N, lung, breast, and prostate cases (Figure 3).

**Table 1:** Absolute and relative (to the maximum dose) MAE and 3D gamma passing rate between denoised and high-statistics reference dose distributions. Values are reported as mean ± standard deviation. The 3 %/2 mm criterion was used in 3D gamma analysis, with a 10 % low-dose threshold.

| Disease site | Region | MAE (Gy (RBE)) | MAE (%) | 3D Gamma (%) |
|---|---|---|---|---|
| H&N | Whole body | 0.19 ± 0.12 | 0.3 ± 0.2 | 96.3 ± 0.9 |
| | CTV | 0.14 ± 0.36 | 0.2 ± 0.5 | 98.3 ± 0.4 |
| | OoIs | 0.19 ± 0.29 | 0.3 ± 0.5 | 97.8 ± 0.8 |
| Lung | Whole body | 0.16 ± 0.13 | 0.3 ± 0.3 | 97.6 ± 0.9 |
| | CTV | 1.7 ± 1.6 | 1.9 ± 2.1 | 96.4 ± 1.0 |
| | OoIs | 0.33 ± 0.31 | 0.6 ± 0.6 | 97.3 ± 1.0 |
| Breast | Whole body | 0.20 ± 0.11 | 0.4 ± 0.2 | 95.8 ± 0.8 |
| | CTV | 1.2 ± 0.3 | 2.7 ± 0.4 | 94.4 ± 1.0 |
| | OoIs | 0.11 ± 0.07 | 0.3 ± 0.2 | 94.3 ± 0.9 |
| Prostate | Whole body | 0.08 ± 0.06 | 0.2 ± 0.1 | 96.2 ± 0.7 |
| | CTV | 1.0 ± 0.7 | 1.7 ± 0.8 | 96.7 ± 1.0 |
| | OoIs | 0.27 ± 0.23 | 0.5 ± 0.3 | 95.8 ± 0.8 |

Abbreviations: CTV for clinical target volume, OoI for organ of interest, MAE for mean absolute error, H&N for head and neck, and RBE for relative biological effectiveness. RBE = 1.1.

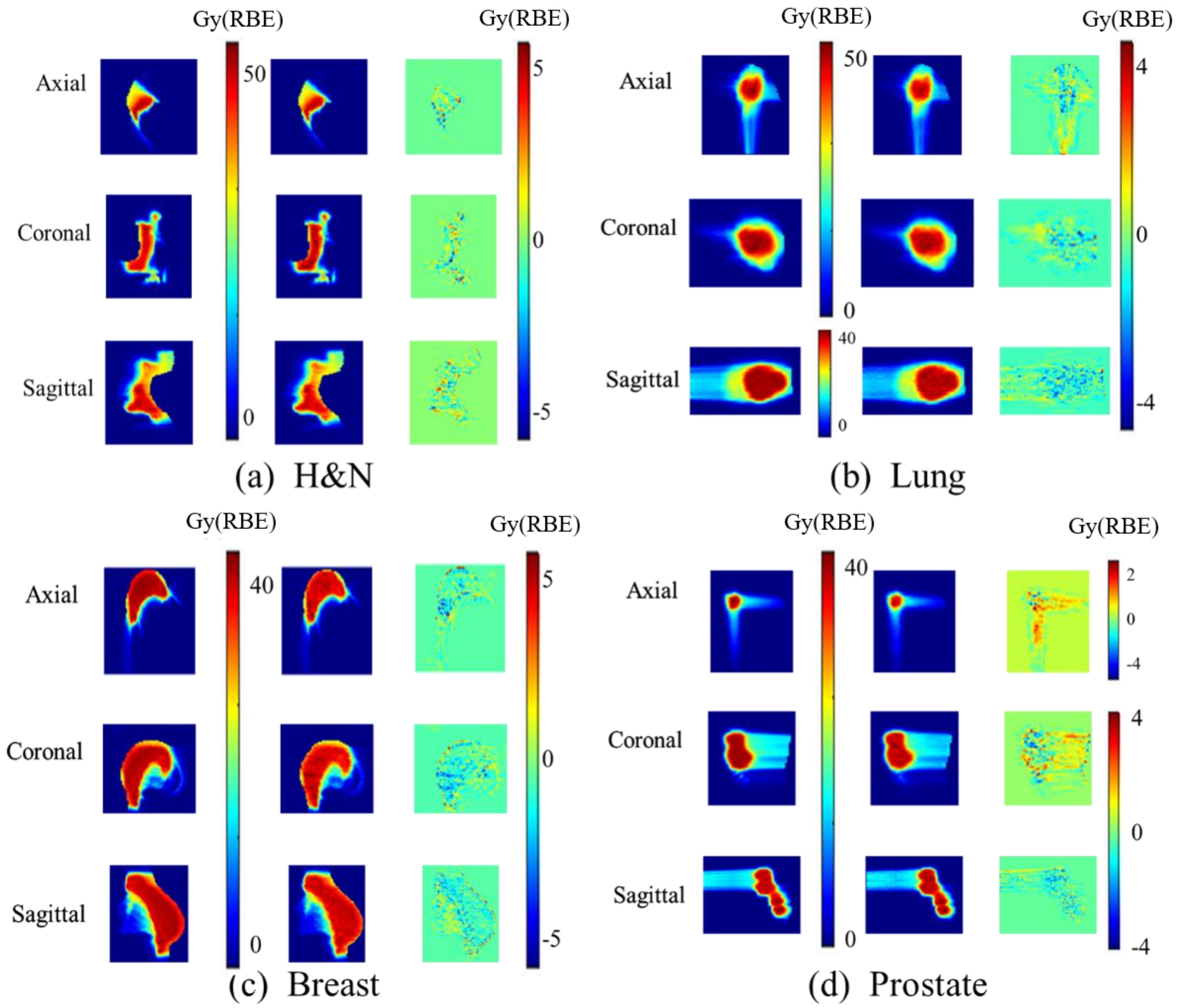

**Figure 2.** Comparison of the high-statistics reference and denoised dose distributions in the axial, coronal, and sagittal planes for representative (a) H&N, (b) lung, (c) breast, and (d) prostate cases. Columns show the reference dose, denoised dose, and dose difference, respectively. The denoised dose distributions closely matched the reference dose distributions across all disease sites.

Under the 3 %/2 mm criterion, mean 3D gamma passing rates ranged from 95.8 ± 0.8 % to 97.6 ± 0.9 % for the whole body, from 94.4 ± 1.0 % to 98.3 ± 0.4 % for the CTV, and from 94.3 ± 0.9 % to 97.8 ± 0.8 % for OoIs across the four evaluated disease sites (Table 1). The 2 %/2 mm sensitivity analysis showed systematically lower passing rates for the whole body, with detailed values and analysis provided in the Supplementary Materials.

Gaussian-filter validation identified σ = 1.88 mm as the filter width associated with the lowest whole-body MAE (0.17 Gy (RBE)). Optimized Gaussian filtering yielded lower whole-body

MAE than the proposed framework in the H&N, lung, and breast cohorts and the same rounded MAE in the prostate cohort, while whole-body 3 %/2 mm gamma passing rates were higher with Gaussian filtering in H&N, lung, and prostate and identical in breast (Table 2). In the H&N cohort, the proposed framework yielded a smaller CTV $D_{98\,\%}$ deviation than Gaussian filtering (0.8 ± 1.0 % vs. 1.2 ± 1.2 %), whereas Gaussian filtering yielded a smaller CTV D2 deviation (0.7 ± 0.4 % vs. 1.3 ± 0.7 %).

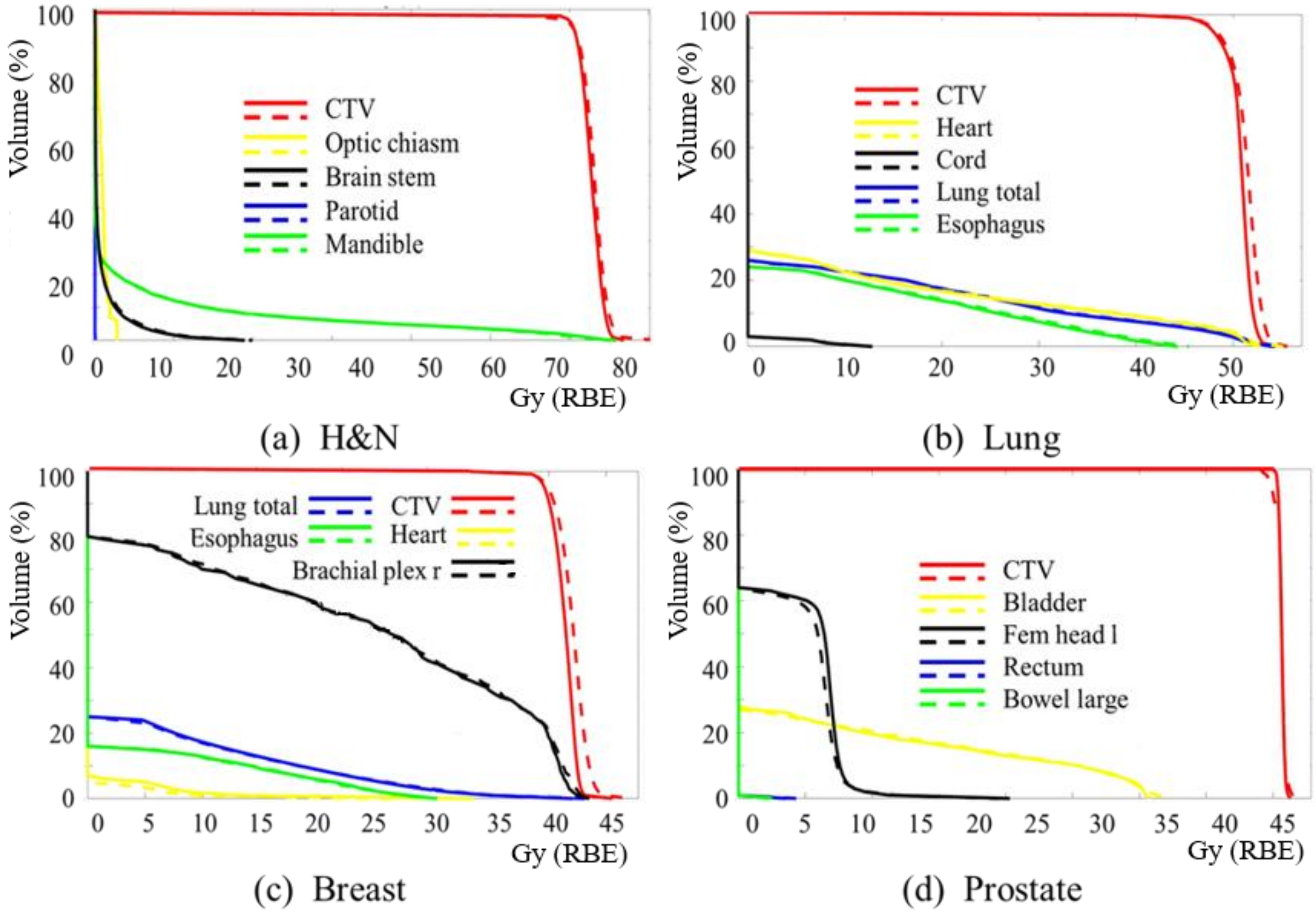

**Figure 3.** Comparison of dose-volume histograms (DVHs) between the high-statistics reference dose (solid lines) and the denoised dose (dashed lines) for representative H&N, lung, breast, and prostate cases.

On our hardware platform, low-statistics MCsquare dose calculations required approximately 1 minute per patient, compared with approximately 10 minutes for high-statistics MCsquare calculations in representative H&N cases. The proposed diffusion-transformer framework required approximately 40–120 s per patient on an NVIDIA GPU, depending on the dose-map dimensions and number of processed patches.

**Table 2.** Low-statistics Monte Carlo dose, optimized Gaussian filter (σ = 1.88 mm), the proposed framework, and the high-statistics reference, evaluated under identical conditions. Values are mean ± standard deviation. CTV $D_{98\,\%}$ and $D_{2\,\%}$ deviations are reported in values relative to the maximum dose.

| Test cohort | Metric | Low-statistics MC | Gaussian σ = 1.88 mm | Proposed framework |
|---|---|---|---|---|
| H&N | Whole-body MAE (Gy (RBE)) | 0.28 ± 0.26 | 0.18 ± 0.12 | 0.19 ± 0.12 |
| | Whole-body 3D Gamma (%) | 85.6 ± 10.7 | 98.6 ± 1.9 | 96.3 ± 0.9 |
| Lung | Whole-body MAE (Gy (RBE)) | 0.15 ± 0.10 | 0.12 ± 0.12 | 0.16 ± 0.14 |
| | Whole-body 3D Gamma (%) | 86.5 ± 9.0 | 99.0 ± 1.1 | 97.6 ± 0.9 |
| Breast | Whole-body MAE (Gy (RBE)) | 0.31 ± 0.20 | 0.18 ± 0.11 | 0.20 ± 0.11 |
| | Whole-body 3D Gamma (%) | 68.4 ± 9.8 | 95.8 ± 2.5 | 95.8 ± 0.8 |
| Prostate | Whole-body MAE (Gy (RBE)) | 0.06 ± 0.06 | 0.06 ± 0.05 | 0.08 ± 0.06 |
| | Whole-body 3D Gamma (%) | 95.3 ± 3.3 | 99.9 ± 0.1 | 96.2 ± 0.7 |
| H&N | CTV $D_{98\,\%}$ deviation (%) | 3.0 ± 3.9 | 1.2 ± 1.2 | 0.8 ± 1.0 |
| | CTV $D_{2\,\%}$ deviation (%) | 4.0 ± 3.3 | 0.7 ± 0.4 | 1.3 ± 0.7 |

Abbreviations: CTV for clinical target volume, MC for Monte Carlo, MAE for mean absolute error, H&N for head and neck, RBE for relative biological effectiveness. RBE = 1.1.

## 4. Discussion

In this study, we developed a diffusion-transformer-based framework for denoising low-statistics MC dose distributions in pencil beam scanning proton therapy. Trained exclusively on H&N data, the model maintained favorable dose/volume metric agreement and cross-site performance in independent lung, breast, and prostate cohorts without fine-tuning or retraining. These findings support the feasibility of cross-site dose denoising using heterogeneous clinical data.

Recent proton-specific deep-learning studies have approached fast dose calculation through either direct dose prediction or MC dose denoising. Direct prediction models have demonstrated promising accuracy but require substantial patient- and plan-specific information. Pirlepesov et al. [43] incorporated anatomical structures, CT information, geometric descriptors, and beam-path features for 3D dose and dose averaged linear energy transfer ($LET_D$) prediction, whereas Wang et al. [44] used stopping-power information, spot energies, MU distributions, and spatial masks for IMPT dose prediction. In contrast, MC denoising starts from a physics-based low-statistics dose distribution that already incorporates patient anatomy and treatment delivery information, thereby constraining the learning task to recovering the corresponding high-statistics dose. Javaid et al. [45] demonstrated this approach across multiple tumor sites using simple convolutional neural network (CNN) and dilated U-Net architectures, while Zhang et al. [46] used a deep ResNet-deconvolution network for fast proton plan verification in online adaptive workflows. The present study extended this line of investigation by using a diffusion-transformer architecture and, importantly, by evaluating cross-site transfer from a model trained exclusively on heterogeneous H&N data to independent lung, breast, and prostate cohorts without retraining. Accordingly, these differences in study design emphasized cross-site generalizability.

The preprocessing pipeline was implemented as a fixed standardized component across disease sites and evaluated as a whole. Because formal retraining-based ablation experiments were not performed, the contributions of individual preprocessing components to the observed performance were not determined. To accommodate limited clinical data and the differences between medical and natural-image domains, we adopted a customized diffusion transformer architecture with eight transformer blocks. This design was intended to balance model expressiveness and computational cost while maintaining robust denoising performance on heterogeneous three-dimensional datasets.

The slightly higher CTV MAE observed in lung, breast, and prostate cases is expected under out-of-distribution evaluation and likely reflects site-specific differences in target geometry and dose heterogeneity rather than fundamentally degraded model performance. Optimized Gaussian filtering achieved whole-body MAE and gamma passing rates comparable to, and in several cohorts better than, those of the proposed framework. These global metrics are influenced by the large volume of low-dose voxels and may therefore be less sensitive to localized differences in target-dose fidelity. In the H&N cohort, the proposed framework yielded a smaller CTV $D_{98\,\%}$ deviation than optimized Gaussian filtering, suggesting better preservation of this target-coverage endpoint. Thus, the learned approach was not uniformly superior to optimized Gaussian filtering but exhibited a different error profile. Formal ablation studies evaluating alternative normalization schemes, preprocessing strategies, and convolutional architectures were not performed in the current work and represent important future directions for further optimization and validation of the framework.

The gamma-analysis results should be interpreted in the context of the native 2.5 mm MC dose-grid resolution. Although interpolation was used during gamma evaluation, it does not increase the intrinsic spatial resolution of the underlying dose distribution; therefore, favorable agreement under the 3 %/2 mm criterion should not be interpreted as evidence of equivalent performance at substantially finer spatial tolerances. Target-region performance may be influenced by site-specific differences in target geometry and dose heterogeneity under out-of-distribution evaluation, as well as by the lack of explicit prioritization of anatomically defined regions in the current framework.

Balancing accuracy and generalizability remain challenging in real-world applications, and generalizable models may perform differently from specialized, task-specific models when applied to previously unseen data [43,44]. For applications requiring high precision, task-specific fine-tuning or domain-aligned preprocessing may further improve reliability, although this would reduce the advantages of a single broadly applicable model.

The proposed framework may support oART workflows where rapid dose recomputation is required, and may also improve planning efficiency in preauthorization workflows. The observed cross-disease-site performance also suggests potential applications in broader clinical and research settings, including retrospective dose analysis and large-scale outcomes research, although further validation is needed before clinical deployment.

Several methodological considerations and limitations should be acknowledged. Structure contours were not provided to the network as predictive or conditioning inputs, which may limit the ability of the model to explicitly prioritize target and OoI regions. Future studies may investigate whether incorporating anatomical structures or other patient-specific clinical information improves performance in clinically important regions. In addition, different normalization references were used during training and inference: the maximum high-statistics dose was used during training, whereas the clinically available prescription dose was used during inference. These two reference values are not necessarily numerically identical; however, changing between them applies a patient-level global scaling factor and preserves voxel ordering, relative dose gradients, and the spatial dose pattern that defines the denoising task. We do not assume that the network is mathematically invariant to this change in scaling. Rather, because all reported testing results were obtained using prescription-dose normalization during inference, the observed reconstruction performance across the H&N, lung, breast, and prostate cohorts provides empirical evidence that the model remained robust to this change in global dose scaling in the evaluated data. Nonetheless, the influence of different normalization factor choices needs a systematic investigation.

Formal ablation studies were also not performed to isolate the contributions of individual preprocessing or architectural components, and a matched 3D U-Net or another optimized convolutional learned baseline was not evaluated. Therefore, the present study does not provide a direct comparison between the DiT architecture and convolutional approaches. Range-shifter cases were not included in the training or testing datasets; because range shifters can introduce additional scattering and range-degradation effects, dedicated evaluation is needed in these scenarios. Finally, the independent disease-site cohorts were relatively small and derived from a single institution, and broader validation will be needed to assess performance across more diverse clinical settings.

In conclusion, this study demonstrated the feasibility of a DiT-based framework for denoising low-statistics MC dose distributions in proton therapy. Trained solely on H&N data, the model maintained cross-site performance in independent H&N, lung, breast, and prostate cohorts without fine-tuning or retraining, while accommodating variations in image dimensions, prescription doses, and beam configurations. These findings support the potential of the framework as a flexible approach to proton dose denoising.

## References

[1] Mohan R, Grosshans D. Proton therapy - present and future. Adv Drug Deliv Rev 2017;109:26-44. https://doi.org/10.1016/j.addr.2016.11.006.

[2] Schild SE, Rule WG, Ashman JB, Vora SA, Keole S, Anand A, et al. Proton beam therapy for locally advanced lung cancer: a review. World J Clin Oncol 2014;5:568-575. https://doi.org/10.5306/wjco.v5.i4.568.

[3] An Y, Shan J, Patel SH, Wong W, Schild SE, Ding X, et al. Robust intensity-modulated proton therapy to reduce high linear energy transfer in organs at risk. Med Phys 2017;44:6138-6147. https://doi.org/10.1002/mp.12610.

[4] Rwigema JCM, Langendijk JA, van der Laan HP, Lukens JN, Swisher-McClure SD, Lin A. A model-based approach to predict short-term toxicity benefits with proton therapy for oropharyngeal cancer. Int J Radiat Oncol Biol Phys 2019;104:553-562. https://doi.org/10.1016/j.ijrobp.2018.12.055.

[5] Blanchard P, Garden AS, Gunn GB, Rosenthal DI, Morrison WH, Hernandez M, et al. Intensity-modulated proton beam therapy (IMPT) versus intensity-modulated photon therapy (IMRT) for patients with oropharynx cancer: a case-matched analysis. Radiother Oncol 2016;120:48-55. https://doi.org/10.1016/j.radonc.2016.05.022.

[6] Frank SJ, Busse PM, Lee JJ, Rosenthal DI, Hernandez M, Swanson DM, et al. Proton versus photon radiotherapy for patients with oropharyngeal cancer in the USA: a multicentre, randomised, open-label, non-inferiority phase 3 trial. Lancet 2026;407:174-184. https://doi.org/10.1016/S0140-6736(25)01962-2.

[7] Chen J, Yang Y, Feng H, Zhang L, Liu Z, Liu T, et al. Robust optimization for spot-scanning proton therapy based on dose-linear-energy-transfer volume constraints. Int J Radiat Oncol Biol Phys 2025;121:1303-1315. https://doi.org/10.1016/j.ijrobp.2024.11.068.

[8] Lomax AJ. Intensity modulated proton therapy and its sensitivity to treatment uncertainties 1: the potential effects of calculational uncertainties. Phys Med Biol 2008;53:1027-1042. https://doi.org/10.1088/0031-9155/53/4/014.

[9] Schneider U, Pedroni E, Lomax A. The calibration of CT Hounsfield units for radiotherapy treatment planning. Phys Med Biol 1996;41:111-124. https://doi.org/10.1088/0031-9155/41/1/009.

[10] Schaffner B, Pedroni E. The precision of proton range calculations in proton radiotherapy treatment planning: experimental verification of the relation between CT-HU and proton stopping power. Phys Med Biol 1998;43:1579-1592. https://doi.org/10.1088/0031-9155/43/6/016.

[11] Liu W, Mohan R, Park P, Liu Z, Li H, Li X, et al. Dosimetric benefits of robust treatment planning for intensity modulated proton therapy for base-of-skull cancers. Pract Radiat Oncol 2014;4:384-391. https://doi.org/10.1016/j.prro.2013.12.001.

[12] Liu W, Zhang X, Li Y, Mohan R. Robust optimization of intensity modulated proton therapy. Med Phys 2012;39:1079-1091. https://doi.org/10.1118/1.3679340.

[13] Shan J, An Y, Bues M, Schild SE, Liu W. Robust optimization in IMPT using quadratic objective functions to account for the minimum MU constraint. Med Phys 2018;45:460-469. https://doi.org/10.1002/mp.12677.

[14] Shan J, Sio TT, Liu C, Schild SE, Bues M, Liu W. A novel and individualized robust optimization method using normalized dose interval volume constraints (NDIVC) for intensity-modulated proton radiotherapy. Med Phys 2019;46:382-393. https://doi.org/10.1002/mp.13276.

[15] Pflugfelder D, Wilkens JJ, Oelfke U. Worst case optimization: a method to account for uncertainties in the optimization of intensity modulated proton therapy. Phys Med Biol 2008;53:1689-1700. https://doi.org/10.1088/0031-9155/53/6/013.

[16] Unkelbach J, Chan TCY, Bortfeld T. Accounting for range uncertainties in the optimization of intensity modulated proton therapy. Phys Med Biol 2007;52:2755-2773. https://doi.org/10.1088/0031-9155/52/10/009.

[17] van de Water S, van Dam I, Schaart DR, Al-Mamgani A, Heijmen BJM, Hoogeman MS. The price of robustness; impact of worst-case optimization on organ-at-risk dose and complication probability in intensity-modulated proton therapy for oropharyngeal cancer patients. Radiother Oncol 2016;120:56-62. https://doi.org/10.1016/j.radonc.2016.04.038.

[18] Inoue T, Widder J, van Dijk LV, Takegawa H, Koizumi M, Takashina M, et al. Limited impact of setup and range uncertainties, breathing motion, and interplay effects in robustly optimized intensity modulated proton therapy for stage III non-small cell lung cancer. Int J Radiat Oncol Biol Phys 2016;96:661-669. https://doi.org/10.1016/j.ijrobp.2016.06.2454.

[19] Shan J, Yang Y, Schild SE, Daniels TB, Wong WW, Fatyga M, et al. Intensity-modulated proton therapy (IMPT) interplay effect evaluation of asymmetric breathing with simultaneous uncertainty considerations in patients with non-small cell lung cancer. Med Phys 2020;47:5428-5440. https://doi.org/10.1002/mp.14491.

[20] Liu C, Schild SE, Chang JY, Liao Z, Korte S, Shen J, et al. Impact of spot size and spacing on the quality of robustly optimized intensity modulated proton therapy plans for lung cancer. Int J Radiat Oncol Biol Phys 2018;101:479-489. https://doi.org/10.1016/j.ijrobp.2018.02.009.

[21] Liu C, Sio TT, Deng W, Shan J, Daniels TB, Rule WG, et al. Small-spot intensity-modulated proton therapy and volumetric-modulated arc therapies for patients with locally advanced non-small-cell lung cancer: a dosimetric comparative study. J Appl Clin Med Phys 2018;19:140-148. https://doi.org/10.1002/acm2.12459.

[22] Liu C, Yu NY, Shan J, Bhangoo RS, Daniels TB, Chiang JS, et al. Technical note: treatment planning system (TPS) approximations matter - comparing intensity-modulated proton therapy (IMPT) plan quality and robustness between a commercial and an in-house developed TPS for nonsmall cell lung cancer (NSCLC). Med Phys 2019;46:4755-4762. https://doi.org/10.1002/mp.13809.

[23] Feng H, Shan J, Ashman JB, Rule WG, Bhangoo RS, Yu NY, et al. Technical note: 4D robust optimization in small spot intensity-modulated proton therapy (IMPT) for distal esophageal carcinoma. Med Phys 2021;48:4636-4647. https://doi.org/10.1002/mp.15003.

[24] Feng H, Sio TT, Rule WG, Bhangoo RS, Lara P, Patrick CL, et al. Beam angle comparison for distal esophageal carcinoma patients treated with intensity-modulated proton therapy. J Appl Clin Med Phys 2020;21:141-152. https://doi.org/10.1002/acm2.13049.

[25] Seco J, Robertson D, Trofimov A, Paganetti H. Breathing interplay effects during proton beam scanning: simulation and statistical analysis. Phys Med Biol 2009;54:N283-N294. https://doi.org/10.1088/0031-9155/54/14/N01.

[26] Meijers A, Knopf AC, Crijns APG, Ubbels JF, Niezink AGH, Langendijk JA, et al. Evaluation of interplay and organ motion effects by means of 4D dose reconstruction and accumulation. Radiother Oncol 2020;150:268-274. https://doi.org/10.1016/j.radonc.2020.07.055.

[27] Holmes J, Feng H, Zhang L, Fix MK, Jiang SB, Liu W. Fast Monte Carlo dose calculation in proton therapy. Phys Med Biol 2024;69:17TR01. https://doi.org/10.1088/1361-6560/ad67a7.

[28] Shan J, Feng H, Morales DH, Patel SH, Wong WW, Fatyga M, et al. Virtual particle Monte Carlo: a new concept to avoid simulating secondary particles in proton therapy dose calculation. Med Phys 2022;49:6666-6683. https://doi.org/10.1002/mp.15913.

[29] Giantsoudi D, Schuemann J, Jia X, Dowdell S, Jiang S, Paganetti H. Validation of a GPU-based Monte Carlo code (gPMC) for proton radiation therapy: clinical cases study. Phys Med Biol 2015;60:2257-2269. https://doi.org/10.1088/0031-9155/60/6/2257.

[30] Fippel M, Soukup M. A Monte Carlo dose calculation algorithm for proton therapy. Med Phys 2004;31:2263-2273. https://doi.org/10.1118/1.1769631.

[31] Souris K, Lee JA, Sterpin E. Fast multipurpose Monte Carlo simulation for proton therapy using multi- and many-core CPU architectures. Med Phys 2016;43:1700-1712. https://doi.org/10.1118/1.4943377.

[32] Hong L, Goitein M, Bucciolini M, Comiskey R, Gottschalk B, Rosenthal S, et al. A pencil beam algorithm for proton dose calculations. Phys Med Biol 1996;41:1305-1330. https://doi.org/10.1088/0031-9155/41/8/005.

[33] Taylor PA, Kry SF, Followill DS. Pencil beam algorithms are unsuitable for proton dose calculations in lung. Int J Radiat Oncol Biol Phys 2017;99(3):750-756. https://doi.org/10.1016/j.ijrobp.2017.06.003.

[34] da Silva J, Ansorge R, Jena R. Fast pencil beam dose calculation for proton therapy using a double-Gaussian beam model. Front Oncol 2015;5:281. https://doi.org/10.3389/fonc.2015.00281.

[35] Younkin JE, Morales DH, Shen J, Shan J, Bues M, Lentz JM, et al. Clinical validation of a ray-casting analytical dose engine for spot scanning proton delivery systems. Technol Cancer Res Treat 2019;18:1533033819887182. https://doi.org/10.1177/1533033819887182.

[36] Isaksson LJ, Mastroleo F, Vincini MG, Marvaso G, Zaffaroni M, Gola M, et al. The emerging role of artificial intelligence in proton therapy: a review. Crit Rev Oncol Hematol 2024;204:104485. https://doi.org/10.1016/j.critrevonc.2024.104485.

[37] Lagedamon V, Leni PE, Gschwind R. Deep learning applied to dose prediction in external radiation therapy: a narrative review. Cancer Radiother 2024;28:402-414. https://doi.org/10.1016/j.canrad.2024.03.005.

[38] Ahn SH, Kim E, Kim C, Cheon W, Kim M, Lee SB, et al. Deep learning method for prediction of patient-specific dose distribution in breast cancer. Radiat Oncol 2021;16:154. https://doi.org/10.1186/s13014-021-01864-9.

[39] Wang Q, Song Y, Hu J, Liang L. DoseNet: an ensemble-based deep learning method for 3D dose prediction in IMRT. In: Proc International Annual Conference on Complex Systems and Intelligent Science (CSIS-IAC); 2023 615–620. https://doi.org/10.1109/CSIS-IAC60628.2023.10364051.

[40] Soomro MH, Alves VGL, Nourzadeh H, Siebers JV. DeepDoseNet: a deep learning model for 3D dose prediction in radiation therapy. arXiv [Preprint]. 2021;arXiv:2111.00077. https://doi.org/10.48550/arXiv.2111.00077.

[41] Gronberg MP, Beadle BM, Garden AS, Skinner H, Gay S, Netherton T, et al. Deep learning-based dose prediction for automated, individualized quality assurance of head and neck radiation therapy plans. Pract Radiat Oncol 2023;13:e282-e291. https://doi.org/10.1016/j.prro.2022.12.003.

[42] Gheshlaghi T, Nabavi S, Shirzadikia S, Moghaddam ME, Rostampour N. A cascade transformer-based model for 3D dose distribution prediction in head and neck cancer radiotherapy. Phys Med Biol 2024;69:045010. https://doi.org/10.1088/1361-6560/ad209a.

[43] Pirlepesov F, Wilson L, Moskvin VP, Breuer A, Parkins F, Lucas JT Jr, et al. Three-dimensional dose and LETD prediction in proton therapy using artificial neural networks. Med Phys 2022;49:7417–7427. https://doi.org/10.1002/mp.16043.

[44] Wang W, Chang Y, Liu Y, Liang Z, Liao Y, Qin B, et al. Feasibility study of fast intensity-modulated proton therapy dose prediction method using deep neural networks for prostate cancer. Med Phys 2022;49:5451–5463. https://doi.org/10.1002/mp.15702

[45] Javaid U, Souris K, Huang S, Lee JA. Denoising proton therapy Monte Carlo dose distributions in multiple tumor sites: A comparative neural networks architecture study. Phys Med 2021;89:93–103. https://doi.org/10.1016/j.ejmp.2021.07.022.

[46] Zhang G, Chen X, Dai J, Men K. A plan verification platform for online adaptive proton therapy using deep learning-based Monte Carlo denoising. Phys Med 2022;103:18–25. https://doi.org/10.1016/j.ejmp.2022.09.018.

[47] Peebles W, Xie S. Scalable diffusion models with transformers. In: Proc IEEE/CVF International Conference on Computer Vision (ICCV); 2023 4195-4205. https://doi.org/10.1109/ICCV51070.2023.00387.

[48] Nguyen D, McBeth R, Sadeghnejad Barkousaraie A, Bohara G, Shen C, Jia X, et al. Incorporating human and learned domain knowledge into training deep neural networks: a differentiable dose-volume histogram and adversarial inspired framework for generating Pareto optimal dose distributions in radiation therapy. Med Phys 2020;47:837-849. https://doi.org/10.1002/mp.13955.